\documentstyle [12pt, epsf] {article}
 
\catcode`@=12
\newcommand{\be}{\begin{equation}}
\newcommand{\ee}{\end{equation}}
\newcommand{\bea}{\begin{eqnarray}}
\newcommand{\eea}{\end{eqnarray}}

\def\lsim{\:\raisebox{-0.5ex}{$\stackrel{\textstyle<}{\sim}$}\:}

\begin{document}
\vspace{0.5in}
\oddsidemargin -.1 in
\newcount\sectionnumber
\sectionnumber=0

\def\lsim{\mathrel{\vcenter{\hbox{$<$}\nointerlineskip\hbox{$\sim$}}}}
\thispagestyle{empty}

\vskip.5truecm
\vspace*{0.5cm}

\begin{center}
{\Large \bf \centerline{ Neutrino mass matrix in the standard parametrization 
 }} {\Large\bf \centerline{with texture two zeros
}} \vspace*{0.5cm} { R. Mohanta$^1$, G. Kranti$^1$, A. K. Giri$^2$}
\vskip0.3cm 
{\it $^1$ School of Physics, University of Hyderabad,
 Hyderabad - 500 046, India}\\
{\it $^2$ Department of Physics, Punjabi University,
 Patiala - 147 002, India}\\

\vskip0.5cm
\bigskip
\begin{abstract}

 We study the texture two zeros neutrino mass matrices using the
standard parametrization for the neutrino mixing matrix. We find that
if the origin of CP violation in the leptonic sector is not due to 
the Dirac-type complex phase of the mixing matrix but because of 
some non-standard
phenomena then some of the possible texture two mass matrices, which
are allowed by standard parametrization, are found to be unsuitable
to accommodate the observed data in the neutrino sector. Furthermore,
incorporating nonzero Dirac phase in our analysis we find that
many of them do not exhibit normal hierarchy. 
\end{abstract}
\end{center}
\thispagestyle{empty}

\newpage


The study of neutrino physics is now one of the hotly pursued areas of
High 
Energy Physics research. The recent experiments on solar, atmospheric,
reactor and accelerator neutrinos  \cite{ref1} have provided 
us an unambiguous 
evidence that neutrinos are massive and lepton flavors are mixed.
Within the standard model neutrinos are strictly massless. Thus the
non-vanishing neutrino mass is the first clear evidence of new
physics beyond the standard model.
Since neutrinos are massive, there will
be flavor mixing in the charged current
interaction of the leptons and a leptonic mixing matrix will appear
analogous to the CKM mixing matrix for the quarks. Thus, the  three
flavor eigenstates of neutrinos ($\nu_e,~ \nu_\mu,~ \nu_\tau$) 
are related to the corresponding
mass eigenstates ($\nu_1,~ \nu_2,~ \nu_3$) by the unitary transformation 

\bea
 \left ( \matrix{
\nu_e       \cr
\nu_\mu \cr
\nu_\tau \cr} 
\right ) \; =\; \left ( \matrix{
V_{e1}      & V_{e2}    & V_{e3} \cr
V_{\mu 1}      & V_{\mu 2}    & V_{\mu 3} \cr
V_{\tau 1}      & V_{\tau 2}    & V_{\tau 3} \cr} 
\right ) \left ( \matrix{
\nu_1 \cr
\nu_2 \cr
\nu_3 \cr} 
\right ) \; ,
\eea
where $V$ is the $3 \times 3 $ unitary
matrix known as PMNS matrix \cite{pmns}, which contains three mixing angles
and three CP violating phases (one Dirac type and two Majorana
type). In general $V$ can be written as $V =UP$, where
$U$ is the unitary matrix analogous to the
quark mixing matrix  and $P$ is a diagonal matrix 
containing two Majorana phases, i.e., $P$= diagonal
$( e^{i \rho}, e^{i \sigma},1)$. The presence of the
leptonic mixing, analogous to that of quark mixing, has
opened up the possibility that CP violation could also be there
in the lepton sector as it exists in the quark sector. In the standard 
parametrization (PDG) the
mixing matrix is given as
\begin{equation}
U \; = \; \left ( \matrix{ c_x c_z & s_x c_z & s_z e^{-i\delta}
\cr -s_x c_y - c_x s_y s_ze^{i\delta} & c_x c_y - s_x s_y
s_ze^{i\delta} & s_y c_z \cr s_x s_y - c_x c_y s_ze^{i\delta}  &
-c_x s_y - s_x c_y s_ze^{i\delta}  & c_y c_z \cr } \right ) \; ,
\end{equation}
where $ \theta_{(x,y,z)}\equiv\theta_{(12,23,13)}$ and
$s_x \equiv \sin\theta_x$, $c_x \equiv \cos\theta_x$, and so
on .

Several analyses have been performed in order to understand the form of 
the neutrino mixing matrix and the pattern of lepton mixing 
appears to be understood. The
2-3 mixing is consistent with maximal, 1-2 mixing
is  large but not maximal, 1-3 mixing is small and
appears to be close to zero. It is thus inferred from the
current experimental data 
that the mixing matrix $U$ involves two large mixing
angles ( $ \theta_{12} \sim 30^\circ$ and $ \theta_{23} \sim 45^\circ$ )
and one small angle ($ \theta_{13} < 12^\circ$)
\cite{ref2}. The best-fit values \cite{ref9} of the mixing angles
 with $2 \sigma $ errors are found to be $\theta_x={34^\circ}_{-2.9^\circ }^{
+3.5^\circ}$, $\theta_y={41.6^\circ}_{-5.7^\circ }^{
+10.4^\circ}$, $\theta_z={5.4^\circ}_{-5.4^\circ }^{
+4.9^\circ}$. On the other hand,
the three CP violating phases $\delta$ (Dirac type), $\rho$ and $\sigma$
(Majorana type) are totally unrestricted.

The study of CP violation in the leptonic sector
is also  very important for
a complete understanding of the neutrino masses and mixing as it
is intimately related to the mixing matrix. Furthermore, there appears
to be no reason why CP violation should not be there in the leptonic sector
keeping in mind the fact that large CP violation has 
already been established in the
quark sector. CP violation
in the leptonic sector occurs in the neutrino oscillation due to the
non vanishing Dirac type phase $\delta $ or due to some symmetry
breaking at very high energy. If one considers the effect
of CP violation is
due to the neutrino flavor-mixing one can then obtain the 
rephasing invariant quantity \cite{jarlskog}
\begin{equation}
J={\rm Im}\left ( U_{\alpha i} U_{\beta j} U_{\alpha j}^* U_{\beta i}^*\right )
=\frac{1}{8} \sin 2 \theta_x \sin 2 \theta_y \sin 2 \theta_z
\cos \theta_z \sin \delta\;.
\end{equation}
Using the current experimental data on the mixing
angles, one thus obtains $J \sim {\cal O}(10^{-2}) \sin \delta $.
Therefore, unless $\delta$ is very small the CP violation
effect could be observable in the long baseline experiments.
However, since CP violation is not observed so far in the lepton sector,
$\delta$ is expected to be negligibly small. In our
analysis, therefore, we would like to see the effect on the neutrino mass
matrix when the Dirac type phase happens to be zero and also when
it is non-zero. 

One of the main objectives of neutrino physics research 
is to identify the form
and the origin of neutrino mass matrix \cite{moh}. Unfortunately,
so far, we have been able to infer only the mass difference squares
for the neutrinos but not the individual
ones apart from the maximal (23), large but not maximal (12) 
and small (13)
mixing angles. Furthermore, there is another important issue which needs to be
settled regarding whether neutrinos respect the normal hierarchy, as in
case of quarks, or to that of inverted hierarchy apart from the very fact
that it is not yet established whether neutrinos are of Dirac type or
of Majorana nature. Dedicated neutrino experiments have already
provided us with the first ever clear evidence of physics beyond the
standard model in the form of non-zero neutrino mass squares. Therefore,
it is a challenging time for the theoretical community to settle down
some of the issues, mentioned above, at the earliest possibility.

Studies based on mass matrices can help us to understand the nature
of neutrinos where one can obtain relations among the individual neutrino
masses and the mixing angles, and those findings in turn, 
alongwith the inputs form the data, 
can guide us to unravel the true nature of the neutrino
mass matrix.  There exist many studies in the
literature regarding the textures in neutrinos as well as in the quark sector.
These studies help us to identify with  the flavor symmetry
 and are also shown to be related to the physics at higher scale, e.g., 
TeV scale physics.  
The spirit of lepton-quark
universality motivates one to assume that the lepton mass
matrices might have the same texture zeros as the quark mass
matrices. Such an assumption is indeed reasonable in some specific
models of grand unified theory (GUT) in which mass matrices of 
leptons and quarks are related to each
other by a new kind of flavor symmetry. It is well known that
the texture two zero quark mass matrices for $M_u$ and $M_d$
are more successful than the corresponding three-zero textures to interpret
 the strong hierarchy of the quark masses and the smallness of flavor mixing
angles. That is why two-zero texture of charged-lepton and neutrino mass 
matrices have been considered as a typical example in some model
buildings. Furthermore, the texture two zero neutrino mass matrices
have more free parameters than texture three zeros, which are
quite suitable to interpret the observed bi-large pattern of lepton flavor
mixing.
Recently, Frampton, Glashow and Marfatia \cite{ref3} have examined the
possibility that  the  lepton mass matrices with texture
two zeros  may describe the current experimental data and obtained seven
acceptable forms. Considering the Fritzsch type
parametrization Xing \cite{ref4} has carried out the investigation 
and obtained
the expressions for  neutrino mass ratios and
calculated the Majorana-type CP-violating
phases for all seven possible textures.

In this paper, we first study the effects of vanishing
Dirac type phase on the texture two neutrino mass matrices. 
We then consider the PDG
standard parametrization for the mixing matrix and obtain 
the ratios of different neutrino
masses. We find  that 
out of the seven possible forms only three are allowed
by the current experimental data, if the Dirac type CP violating phase
happens to be zero. We thereafter study the
case of non-zero Dirac phase and obtain interesting results.

In the flavor basis, where the charged lepton mass matrix is
diagonal, the neutrino mass matrix can be written as
\begin{equation}
M \; =\; V \left ( \matrix{
m_1 & 0 & 0 \cr
0 & m_2 & 0 \cr
0 & 0 & m_3 \cr} \right ) V^T =  U \left ( \matrix{
\lambda_1 & 0 & 0 \cr
0 & \lambda_2 & 0 \cr
0 & 0 & \lambda_3 \cr} \right ) U^T \; ,
\end{equation}
where $m_i$ (for $i=1,2,3$) denote the real and positive neutrino masses,
and   $\lambda_i$ are the complex neutrino mass eigenvalues
which include the two Majorana-type CP-violating phases
\begin{equation}
\lambda_1 \; =\; m_1 e^{2i\rho} \; , ~~~
\lambda_2 \; =\; m_2 e^{2i\sigma} \; , ~~~
\lambda_3 = m_3 \; .\label{eq1}
\end{equation}
Since $M$ is symmetric with two texture zeros one can immediately
obtain the constraint relations \cite{ref4}
\begin{equation}
\sum_{i=1}^{3} \left (U_{ai} U_{bi} \lambda_i \right ) = 0 \; , ~~~~~~
\sum_{i=1}^{3} \left (U_{\alpha i} U_{\beta i} \lambda_i \right ) = 0 \; ,
\end{equation}
where each of the four subscripts run over
$e$, $\mu$ and $\tau$, but $(\alpha, \beta) \neq (a, b)$.
Solution of Eq. (6) yields
\begin{equation}
\frac{\lambda_1}{\lambda_3} \; =\;
\frac{U_{a3} U_{b3} U_{\alpha 2} U_{\beta 2} - U_{a2} U_{b2} U_{\alpha 3}
U_{\beta 3}}{U_{a2} U_{b2} U_{\alpha 1} U_{\beta 1} - U_{a1} U_{b1}
U_{\alpha 2} U_{\beta 2}} \; ,\label{eq2}
\end{equation}
and
\begin{equation}
\frac{\lambda_2}{\lambda_3} \; =\;
\frac{U_{a1} U_{b1} U_{\alpha 3} U_{\beta 3} - U_{a3} U_{b3} U_{\alpha 1}
U_{\beta 1}}{U_{a2} U_{b2} U_{\alpha 1} U_{\beta 1} - U_{a1} U_{b1}
U_{\alpha 2} U_{\beta 2}} \;.\label{eq3}
\end{equation}
Now comparing Eqs. (\ref{eq2}) and (\ref{eq3}) with Eq. (\ref{eq1}), 
one can obtain the expressions of
neutrino mass ratios as follows:
\begin{eqnarray}
\frac{m_1}{m_3} & = & \left |
\frac{U_{a3} U_{b3} U_{\alpha 2} U_{\beta 2} - U_{a2} U_{b2} U_{\alpha 3}
U_{\beta 3}}{U_{a2} U_{b2} U_{\alpha 1} U_{\beta 1} - U_{a1} U_{b1}
U_{\alpha 2} U_{\beta 2}} \right | \; ,
\nonumber \\ \nonumber \\
\frac{m_2}{m_3} & = & \left |
\frac{U_{a1} U_{b1} U_{\alpha 3} U_{\beta 3} - U_{a3} U_{b3} U_{\alpha 1}
U_{\beta 1}}{U_{a2} U_{b2} U_{\alpha 1} U_{\beta 1} - U_{a1} U_{b1}
U_{\alpha 2} U_{\beta 2}} \right | \;,
\end{eqnarray}
and the two Majorana phases are found to be
\begin{eqnarray}
\rho & = &
\frac{1}{2} \arg \left [ \frac{U_{a3} U_{b3} U_{\alpha 2} U_{\beta 2}
- U_{a2} U_{b2} U_{\alpha 3}
U_{\beta 3}}{U_{a2} U_{b2} U_{\alpha 1} U_{\beta 1} - U_{a1} U_{b1}
U_{\alpha 2} U_{\beta 2}} \right ] \; ,
\nonumber \\ \nonumber \\
\sigma & = & \frac{1}{2} \arg \left [
\frac{U_{a1} U_{b1} U_{\alpha 3} U_{\beta 3} - U_{a3} U_{b3} U_{\alpha 1}
U_{\beta 1}}{U_{a2} U_{b2} U_{\alpha 1} U_{\beta 1} - U_{a1} U_{b1}
U_{\alpha 2} U_{\beta 2}} \right ] \; .
\end{eqnarray}

Furthermore, the ratio of the mass square differences, which is
basically the ratio of solar and atmospheric  mass square differences
is give  as \cite{ref9}
\begin{equation}
R_\nu \; \equiv \; \left | \frac{m^2_2 - m^2_1} {m^2_3 - m^2_2}
\right | =\frac{\Delta m^2_{\rm sun}}{\Delta m^2_{\rm atm}}\;  \; 
\approx \; 0.033\pm 0.008 \; .
\end{equation}

The Majorana nature of the neutrinos allows us to probe
one element of the mass matrix directly. The decay width for the
nutrinoless double $\beta $ decay, i.e., $(A,Z) \to (A, Z+2)+2e^-$,
 a second order weak process, is proportional to the
effective mass given as
\begin{equation}
| M_{ee} | \; = \; m_3 \left | \frac{m_1}{m_3} U^2_{e1} e^{2i\rho}
+ \frac{m_2}{m_3} U^2_{e2} e^{2i\sigma} + U^2_{e3} \right | \; .
\end{equation}
Thus, the $ee$ element of the mass matrix $M$ can be directly 
obtained from the
experiment.

\vspace{0.25cm}

Now we evaluate the above quantities
using  the standard parametrization and with Dirac type phase as zero
for the flavor mixing
matrix $U$:
\begin{equation}
U \; = \; \left ( \matrix{ c_x c_z & s_x c_z & s_z \cr -s_x c_y -
c_x s_y s_z & c_x c_y - s_x s_y s_z & s_y c_z \cr s_x s_y - c_x
c_y s_z  & -c_x s_y - s_x c_y s_z & c_y c_z \cr } \right ) \;.
\end{equation}

\underline{Pattern $\rm A_1:$} ~ $M_{ee} = M_{e\mu} = 0$ (i.e.,
$a=b=e$; $\alpha = e$ and $\beta =\mu$). By use of Eqs. (7)--(12), 
we obtain the mass ratios as 
\begin{eqnarray}
\frac{\lambda_1}{\lambda_3} & = & \frac{s_z}{c_z^2}\left [ t_x t_y
- s_z \right ] \; ,
\nonumber \\
\frac{\lambda_2}{\lambda_3} & = & - \frac{s_z}{c_z^2}\left [
\frac{t_y}{t_x}+ s_z \right ]  \; .
\end{eqnarray}
Since the 1-3 mixing angle ($\theta_z$) is very small, it is 
appropriate to take the limit $s_z^2 <<1$ and $c_z^2 \to 1$. In
this limit, one can explicitly obtain the different mass ratios and the
Majorana type phases as
\begin{eqnarray}
&& \frac{m_1}{m_3} \; \approx \; s_z t_x t_y  \; , ~~~
\frac{m_2}{m_3} \; \approx \; s_z \frac{t_y}{t_x}  \; ; ~~~ \rho
\; \approx \; 0 \; , ~~~ \sigma \; \approx \;  \pm \frac{\pi}{2}
\; ;
\nonumber \\
&& R_\nu \; \approx \; \frac{t^2_y}{t^2_x} \left | 1 - t^4_x
\right | s^2_z \; , ~~~ |M_{ee}| \; = \; m_3 s_z^2  \;.
\end{eqnarray}
Now using the central values of the mixing angles 
from \cite{ref9} i.e., $\theta_x=34^\circ$, $\theta_y=42^\circ$
and $\theta_z=5^\circ$, we obtain the values of the mass ratios as
\begin{eqnarray}
\frac{m_1}{m_3} \approx 0.053\;,~~~~\frac{m_2}{m_3}\approx 0.116\;,~~~~~
R_\nu \approx 0.01\;,~~~~|M_{ee}|/m_3 = 0.0076\;.\label{a1}
\end{eqnarray}
Thus, from Eq. (\ref{a1}), it can be seen that this pattern
of mass matrix corresponds to the normal hierarchy case i.e.,
$m_1 < m_2 < m_3$. The ratio of mass square differences 
$R_\nu $ is found to be ${\cal O}(10^{-2})$, which is 
consistent with the ratio of solar to atmospheric 
squared mass differences.

\vspace{0.25cm}

\underline{Pattern $\rm A_2$:} ~ $M_{ee} = M_{e\tau} = 0$ (i.e.,
$a=b=e$; $\alpha =e$ and $\beta =\tau$). In this case the mass ratios 
are given as
\begin{eqnarray}
\frac{\lambda_1}{\lambda_3} & = &
 - \frac{s_z}{c_z^2} \left [\frac{t_x}{t_y} +
s_z \right ] \; ,
\nonumber \\
\frac{\lambda_2}{\lambda_3} & = & \frac{s_z}{ c_z^2} \left
[\frac{1}{t_x t_y} - s_z \right ]\;
 \; .
\end{eqnarray}
As done for A1, in the lowest-order approximation, 
we explicitly obtain
\begin{eqnarray}
&& \frac{m_1}{m_3} \; \approx \; \frac{t_x}{t_y} s_z \; , ~~~
\frac{m_2}{m_3} \; \approx \; \frac{1}{t_x t_y} s_z \; ; ~~~ \rho
\; \approx \;  \pm\frac{\pi}{2} \; , ~~~ \sigma \; \approx \; 0 \;
;
\nonumber \\
&& R_\nu \; \approx \; \frac{1}{t^2_x t^2_y} \left | 1 - t^4_x
\right | s^2_z \; , ~~~ |M_{ee}| \; = \; m_3 s_z^2 \; .
\end{eqnarray}
Again using the values of the mixing angles, as given above,
we obtain the numerical values of different mass ratios as
\begin{eqnarray}
\frac{m_1}{m_3} \approx 0.065\;,~~~~\frac{m_2}{m_3}\approx 0.143\;,~~~~~
R_\nu \approx 0.016\;,~~~~|M_{ee}|/m_3=~ 0.0076\;.
\end{eqnarray}
This pattern gives results almost similar to pattern A1 and
corresponds to normal hierarchy nature of neutrino masses. 
It is very difficult to differentiate between these two
patterns from the experimental data

 \vspace{0.25cm}

\underline{Pattern $\rm B_1$:} ~ $M_{\mu\mu} = M_{e\tau} = 0$
(i.e., $a=b=\mu$; $\alpha =e$ and $\beta =\tau$). Here, we obtain
\begin{eqnarray}
\frac{\lambda_1}{\lambda_3} & = & \frac{s_x s_y c_x (2 s_z^2
c_y^2-s_y^2 c_z^2) - c_y s_z(c_x^2 c_y^2+s_x^2 s_y^2)}{s_x s_y c_x
c_y^2+c_y^3 s_z(s_x^2-c_x^2)+s_x s_y s_z^2c_x(1+c_y^2)} \; ,
\nonumber \\
\frac{\lambda_2}{\lambda_3} & = & \frac{s_x s_y c_x (2 s_z^2
c_y^2-s_y^2 c_z^2) + c_y s_z(s_x^2 c_y^2+s_y^2 c_x^2)}{s_x c_x s_y
c_y^2+c_y^3 s_z(s_x^2-c_x^2)+s_x c_x s_y s_z^2(1+c_y^2)} \; ,
\end{eqnarray}
The smallness of $s^2_z$ allows us to make a similar analytical
approximation as before. To lowest order, we find
\begin{eqnarray}
&& \frac{m_1}{m_3} \; \approx \; \frac{m_2}{m_3} \; \approx \;
t^2_y \; ; ~~~ \rho \; \approx \; \sigma \; \approx  \pm
\frac{\pi}{2} \;;~~~~~~~~
 R_\nu \; \approx \; \frac{1 + t^2_x}{t_x}~
 t_{2y} ~ s_z \; ,
\nonumber\\
&&|M_{ee}| \;  \approx  \; m_3 \left [ t^2_y +\frac{(1-t_x^2)}{t_x
t_y}~s_z\right] \; ,
\end{eqnarray}
where $t_{2y} \equiv \tan 2\theta_y$. It should be noted that
$m_1$ and $m_2$ are not exactly degenerate and their difference is
given as
\begin{eqnarray}
&& \frac{m_2}{m_3}-\frac{m_1}{m_3} \; \approx \; \frac{4 s_z
}{s_{2y} s_{2x}} \;.
\end{eqnarray}
Again using the values of the mixing angles we obtain
\begin{eqnarray}
\frac{m_1}{m_3} \approx \frac{m_2}{m_3}
\approx 0.81\;,~~~~~
R_\nu \approx 1.79\;,~~~~|M_{ee}|/m_3\approx ~ 0.89.
\end{eqnarray}
As seen from above equation, this pattern corresponds
to $ m_1 \approx m_2 \approx m_3$ and the ratio of mass
difference square $R_\nu$  is found to be ${\cal O}(1)$.
If we vary the mixing angles $\theta_x$ and $\theta_y$
within their $2 \sigma $ range (i.e., $31.1^\circ \leq \theta_x 
\leq 37.5^\circ $ and  $35.9^\circ \leq \theta_y \leq 52^\circ $)
and  $\theta_z$ between $(1^\circ -12^\circ)$, the allowed region 
in the $R_\nu - s_z$ parameter plane is shown in Figure-1. 
It should be noted here that we have ignored the case of $\theta_z$=0,
since it will give $R_\nu$=0 and we know from the data that $R_\nu$ is
non-zero. Second, there is no compelling reason (keeping
in mind the quark mixing angles) so as to take it to be zero, although
in the literature one can find the explanation that it could be zero 
under certain symmetry condition
(e.g., $\mu$-$\tau$ symmetry). But again this symmetry has to be broken
to incorporate CP violation in the neutrino sector. 
The minimum value of $R_\nu$ is found to be
0.15 for this case which is nearly five times larger 
the observed $R_\nu$ value.
Thus, this pattern is ruled out by the current experimental data.

\begin{figure}[htb]
   \centerline{\epsfysize 5.5 truein \epsfbox{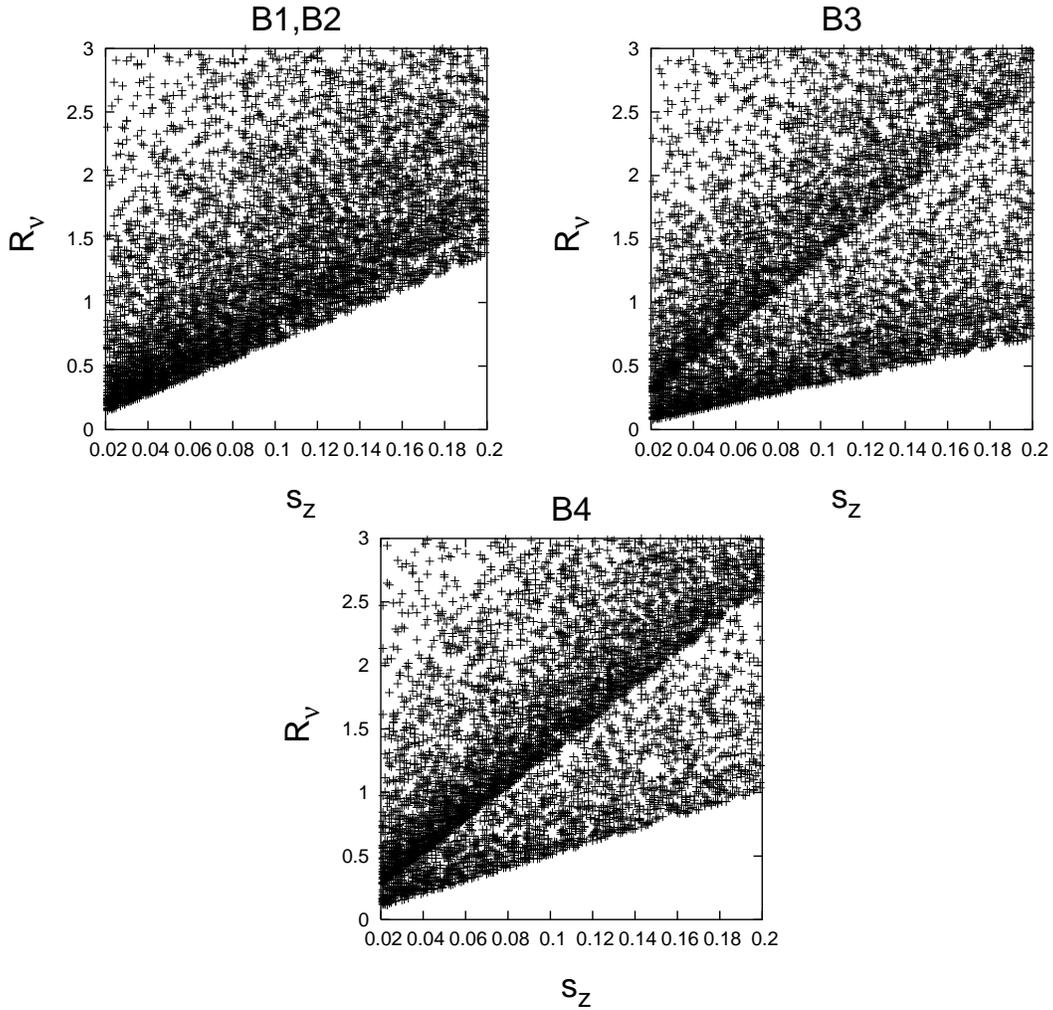}}
 \caption{
  The  allowed region in the $R_\nu - s_z$ parameter plane.}
  \end{figure}

\vspace{0.25cm}

\underline{Pattern $\rm B_2$:} ~ $M_{e\mu} = M_{\tau\tau} = 0$
(i.e., $a=b=\tau$; $\alpha =e$ and $\beta =\mu$). Here, the mass ratios are
given as
\begin{eqnarray}
\frac{\lambda_1}{\lambda_3} & = & \frac{  s_x c_x c_y (2 s_z^2
s_y^2-c_y^2 c_z^2) + s_y s_z(s_x^2 c_y^2+s_y^2 c_x^2) } { s_x c_x
c_y s_y^2-(s_x^2 -c_x^2)s_y^3 s_z+s_x c_x c_y s_z^2(1+s_y^2)} \; ,
\nonumber \\
\nonumber\\
\frac{\lambda_2}{\lambda_3} & = & \frac{ c_x s_x c_y (2 s_z^2
s_y^2-c_y^2 c_z^2)-s_y s_z(s_x^2 s_y^2+c_x^2 c_y^2)}{s_x c_x c_y
s_y^2-(s_x^2 -c_x^2)s_y^3 s_z+s_x c_x c_y s_z^2(1+s_y^2)} \; .
\end{eqnarray}
In the lowest-order approximation, we explicitly obtain
\begin{eqnarray}
&& \frac{m_1}{m_3} \; \approx \; \frac{m_2}{m_3} \; \approx \;
\frac{1}{t_y^2} \; ; ~~~ \rho \; \approx \; \sigma \; \approx  \pm
\frac{\pi}{2} \; ;
~~~~~~~ R_\nu \; \approx \; \frac{1 + t^2_x}{t_x}~
 t_{2y}  ~  s_z \; ,
\nonumber\\
&&|M_{ee}| \;  \approx  \; m_3 \left [ \frac{1}{t^2_y} - \frac{
s_z t_y} {t_x}~(1-t_x^2) \right ] \; ,~~~~~~
\frac{m_1}{m_3}-\frac{m_2}{m_3} \; \approx \; \frac{4 s_z
}{s_{2y} s_{2x}} \;.
\end{eqnarray}
Numerically they are found to be
\begin{eqnarray}
\frac{m_1}{m_3} \approx \frac{m_2}{m_3}\approx 1.23\;,~~~~~
R_\nu \approx 1.79\;,~~~~|M_{ee}|/m_3 \approx ~ 1.79\;.
\end{eqnarray}
This pattern is also similar to B1 and hence not acceptable
by the current data.
\vspace{0.25cm}

\underline{Pattern $\rm B_3$:} ~ $M_{\mu\mu} = M_{e\mu} = 0$
(i.e., $a=b=\mu$; $\alpha =e$ and $\beta =\mu$). We obtain
\begin{eqnarray}
\frac{\lambda_1}{\lambda_3} & = & \; - \frac{s_y}{c_y} \left (
\frac{s_x s_y-c_x c_y s_z}{s_x c_y+c_x s_y s_z}\right )\; ,
\nonumber \\
\nonumber\\
\frac{\lambda_2}{\lambda_3} & =& \; - \frac{s_y}{c_y} ~\left ( \frac{c_x
s_y + s_x c_y s_z }{c_x c_y-s_x s_y s_z} \right )\; .
\end{eqnarray}
The approximate expressions for the neutrino mass ratios, the
Majorana phases and the observables $R_\nu$ and $|M_{ee}|$ turn out
to be
\begin{eqnarray}
&& \frac{m_1}{m_3} \; \approx \; \frac{m_2}{m_3} \; \approx \;
t^2_y \; ; ~~~ \rho \; \approx \; \sigma \; \approx  \pm
\frac{\pi}{2} \; ;
~~~~~~~
 R_\nu \; \approx \; \frac{1 + t^2_x}{t_x}
 ~t^2_y|t_{2y}|~s_z \; ,
\nonumber\\
&&|M_{ee}| \;  \approx  \; m_3 \left [ t^2_y - \frac{
(1-t^2_x)}{t_x}~t_y s_z \right]\; ,~~~~~~
\frac{m_1}{m_3}-\frac{m_2}{m_3} \; \approx \; \frac{4 s_z t^2_y
}{s_{2y} s_{2x}} \;.
\end{eqnarray}
Substituting the values of the mixing angles, we obtain the
numerical values as
\begin{eqnarray}
\frac{m_1}{m_3} \approx \frac{m_2}{m_3}\approx 0.81\;,~~~~~
R_\nu \approx 1.45~~~~{\rm and}~~~~|M_{ee}|/m_3 = 0.75\;.
\end{eqnarray}
B3 is also similar to B1 with $R_\nu ={\cal O}(1)$. 
The allowed region in the parameter space for this case 
(where $\theta_x$ and
$\theta_y$ are varied within their $2 \sigma$ ranges)
 is also shown 
in figure-1 and the minimum $R_\nu$ value obtainable in this case
is 0.08, which is nearly two times greater than the observed $R_\nu$.
Thus, this form is also not acceptable by the current experimental data.
\vspace{0.25cm}

\underline{Pattern $\rm B_4$:} ~ $M_{\tau\tau} = M_{e\tau} = 0$
(i.e., $a=b=\tau$; $\alpha =e$ and $\beta =\tau$). We obtain
\begin{eqnarray}
\frac{\lambda_1}{\lambda_3} & = & \; - \frac{c_y}{s_y} \left (
\frac{s_x c_y+c_x s_y s_z }{s_x s_y-c_x c_y s_z}\right )\; ,
\nonumber \\
\nonumber\\
\frac{\lambda_2}{\lambda_3} & =& \; - \frac{c_y}{s_y} \left (
\frac{c_x c_y - s_x s_y s_z }{c_x s_y+s_x c_y s_z }\right )\; .
\end{eqnarray}
To lowest order, we get the following approximate results:
\begin{eqnarray}
&& \frac{m_1}{m_3} \; \approx \; \frac{m_2}{m_3} \; \approx \;
\frac{1}{t^2_y} \; ; ~~~ \rho \; \approx \; \sigma \; \approx  \pm
\frac{\pi}{2} \; ;~~~~~~
 R_\nu \; \approx \; \frac{1 + t^2_x}{t_x t^2_y}~
 |t_{2y}|  s_z \; ,
\nonumber\\
&&|M_{ee}| \;  \approx  \; m_3 \left [ \frac{1}{t^2_y} +
\frac{1-t^2_x}{t_x t_y}~s_z \right]\; ,
~~~~~~~\frac{m_1}{m_3}-\frac{m_2}{m_3} \; \approx \; \frac{4 s_z
}{t^2_y s_{2y} s_{2x}} \;.
\end{eqnarray}
Using the central values of the mixing angles we obtain
\begin{eqnarray}
\frac{m_1}{m_3} \approx \frac{m_2}{m_3}\approx 1.23\;,~~~~~
R_\nu \approx 2.2~~ ~~{\rm and}~~~~|M_{ee}|/m_3\approx 1.31\;.
\end{eqnarray}
This pattern is also similar to the earlier B's and 
allowed region in the parameter space is also shown in Figure-1. In this
case the minimum
$R_\nu$ value is found to be 0.1, and hence,
 this patten is also unacceptable.
\vspace{0.25cm}

\underline{Pattern $\rm C$:} ~ $M_{\mu\mu} = M_{\tau\tau} = 0$
(i.e., $a=b=\mu$; $\alpha =\beta =\tau$). We obtain
\begin{eqnarray}
\frac{\lambda_1}{\lambda_3} & = & \; - \frac{c_x c_z^2}{s_z} \; .
\frac{c_x(s_y^2-c_y^2)+2 s_x s_y c_y s_z}{2 s_x c_x s_y
c_y-s_z(c_x^2-s_x^2)(c_y^2-s_y^2)+2 s_x c_x s_y c_y s_z^2 }\; ,
\nonumber \\
\nonumber\\
\frac{\lambda_2}{\lambda_3} & = &  \frac{s_x c_z^2}{s_z} \; .
\frac{s_x(s_y^2-c_y^2)-2 c_x s_y c_y s_z}{2 s_x c_x s_y
c_y-(c_x^2-s_x^2)(c_y^2-s_y^2)s_z +2 s_x c_x s_y c_y s_z^2}\;.
\end{eqnarray}
Assuming $s_z^2 << 1$ , we obtain
\begin{eqnarray}
&& \frac{m_1}{m_3} \; \approx \; \left ({1-\frac{1}{t_x t_{2y}
s_z}}\right )\;,~~~~~
 \frac{m_2}{m_3} \; \approx \; \left ({1+\frac{ t_x }{ t_{2y}
s_z}}\right )\;,~~~~~~~\rho =\sigma = \pm \pi/2\;,
\nonumber \\
&& R_\nu \; \approx \; \frac{1+t_x^2}{t_x^2}~
\left|\frac{2}{t_{2x}}~.\frac{1-t_{2x} t_{2y} s_z}{t_x+2 s_z
t_{2y}}\right|\;,
~~~~~~~~
|M_{ee}| \;  \approx  \; m_3\left[1-\frac{2}{ t_{2x} t_{2y}
s_z}\right]\;.
\end{eqnarray}
We find that, in this case $R_\nu $ is very sensitive to the value
of $\theta_z$ and  is found to be ${\cal O}(1)$ for 
$\theta_z=5^\circ$ (the other mixing angles being same as before).
However, $R_\nu$ is found to be acceptable-one
for $\theta_z\approx 2.5^\circ$.
The numerical values of the mass ratios are given as
\begin{eqnarray}
\frac{m_1}{m_3} \approx 2.57\;,~~~~\frac{m_2}{m_3}\approx 2.62\;,~~~~~
R_\nu \approx 0.05 \;,~~~~|M_{ee}|/m_3 \approx~ 0.95\;.
\end{eqnarray}
This pattern corresponds to the mass structure $m_1\approx m_2 >m_3$. 
Thus, although this patten is allowed by the current experimental
data, it is very sensitive for model building.

Thus, we find that out of the seven possibilities, which are allowed 
when $\delta \neq 0 $, only three of them are found to be 
allowed for $\delta=0$, which is our main result.
The order of magnitude of the 
mass matrix $M$ (4) for the three acceptable forms for $\delta=0$ are
given in Table-1. 

Now we will repeat the above procedure for
$\delta \neq 0$. Although the same has been done by Xing for the 
Fritzsch type parametrization, we will redo
it here for the exact standard parametrization (PDG parametrization) 
and would like to see
if there could be any differences that can arise due to the 
difference in the position of the Dirac phase. 

\underline{Pattern $\rm A_1:$} ~ $M_{ee} = M_{e\mu} = 0$ (i.e.,
$a=b=e$; $\alpha = e$ and $\beta =\mu$). We obtain
\begin{eqnarray}
\frac{\lambda_1}{\lambda_3} & = & \frac{s_z}{c_z^2}\left [ t_x t_y
e^{i\delta}- s_z \right ] e^{-2i\delta} \; ,
\nonumber \\
\frac{\lambda_2}{\lambda_3} & = & - \frac{s_z}{c_z^2}\left [
\frac{t_y}{t_x}e^{i \delta}+ s_z \right ] e^{-2 i \delta} \; .
\end{eqnarray}
Again in the lowest order approximation we
obtain the mass ratios as 
\begin{eqnarray}
&& \frac{m_1}{m_3} \; \approx \; s_z t_x t_y  \; , ~~~
\frac{m_2}{m_3} \; \approx \; s_z \frac{t_y}{t_x}  \; ; ~~~ \rho
\; \approx \; -\frac{\delta}{2} \; , ~~~ \sigma \; \approx \;
-\frac{\delta}{2} \pm \frac{\pi}{2} \; ;
\nonumber \\
&& R_\nu \; \approx \; \frac{t^2_y}{t^2_x} \left | 1 - t^4_x
\right | s^2_z \; , ~~~ |M_{ee}| \; = \; m_3 s_z^2  \;.
\end{eqnarray}
The numerical values of the mass ratios are same as that of the pattern A1
of without Dirac phase. So we are not presenting them here. Only the
Majorana phases differ in these two cases.

\vspace{0.25cm}

\underline{Pattern $\rm A_2$:} ~ $M_{ee} = M_{e\tau} = 0$ (i.e.,
$a=b=e$; $\alpha =e$ and $\beta =\tau$). We obtain
\begin{eqnarray}
\frac{\lambda_1}{\lambda_3} & = &
 - \frac{s_z}{c_z^2} \left [\frac{t_x}{t_y} e^{i \delta}+
s_z \right ] e^{-2 i \delta}\; ,
\nonumber \\
\frac{\lambda_2}{\lambda_3} & = & \frac{s_z}{ c_z^2} \left
[\frac{1}{t_x t_y} e^{i\delta}- s_z \right ] e^{-2i\delta}
 \; ,
\end{eqnarray}
which in the limit $s_z^2 <<1$, reduces to
\begin{eqnarray}
&& \frac{m_1}{m_3} \; \approx \; \frac{t_x}{t_y} s_z \; , ~~~
\frac{m_2}{m_3} \; \approx \; \frac{1}{t_x t_y} s_z \; ; ~~~ \rho
\; \approx \;  -\frac{\delta}{2}\pm \frac{\pi}{2} \; , ~~~ \sigma
\; \approx \; -\frac{\delta}{2} \; ;
\nonumber \\
&& R_\nu \; \approx \; \frac{1}{t^2_x t^2_y} \left | 1 - t^4_x
\right | s^2_z \; , ~~~ |M_{ee}| \; = \; m_3 s_z^2 \; .
\end{eqnarray}
The ratio of mass parameters are also same as that of with
$\delta=0$ case.
 \vspace{0.25cm}

\underline{Pattern $\rm B_1$:} ~ $M_{\mu\mu} = M_{e\tau} = 0$
(i.e., $a=b=\mu$; $\alpha =e$ and $\beta =\tau$). We obtain
\begin{eqnarray}
\frac{\lambda_1}{\lambda_3} & = & \frac{s_x c_x s_y (2 s_z^2
c_y^2-s_y^2 c_z^2) - c_y s_z(c_x^2 c_y^2 e^{-i \delta }+s_x^2
s_y^2 e^{i \delta})}{s_x c_x s_y c_y^2+c_y^3 s_z(s_x^2-c_x^2)e^{i
\delta}+s_x c_x s_y s_z^2(1+c_y^2)e^{2 i \delta}} \; ,
\nonumber \\
\nonumber\\
\frac{\lambda_2}{\lambda_3} & = & \frac{s_x c_x s_y (2 s_z^2
c_y^2-s_y^2 c_z^2) + c_y s_z(c_x^2 s_y^2 e^{i \delta }+s_x^2 c_y^2
e^{-i \delta})}{s_x c_x s_y c_y^2+c_y^3 s_z(s_x^2-c_x^2)e^{i
\delta}+s_x c_x s_y s_z^2(1+c_y^2)e^{2 i \delta}} \;.
\end{eqnarray}
 To the lowest order, we find
\begin{eqnarray}
&& \frac{m_1}{m_3} \; \approx \; \frac{m_2}{m_3} \; \approx \;
t^2_y \; ; ~~~ \rho \; \approx \; \sigma \; \approx  \pm
\frac{\pi}{2} \; ;
~~~~~~ R_\nu \; \approx \; \frac{1 + t^2_x}{t_x}~
 |t_{2y} ~ c_{\delta}| ~  s_z \; ,
\nonumber\\
&&|M_{ee}| \;  \approx  \; m_3 \left [ t^2_y + \frac{c_\delta
s_z}{t_x t_y} \left ( (1-t_x^2)(1+t_y^2)\right ) \right ] \;,\label{b11}
\end{eqnarray}
where $c_\delta \equiv
\cos\delta$. Also,
\begin{eqnarray}
&& \frac{m_1}{m_3}-\frac{m_2}{m_3} \; \approx \; \frac{4 s_z
c_{\delta}}{s_{2y} s_{2x}} \; ,~~~ \sigma - \rho \; \approx \;
\frac{2 s_z s_{\delta}}{s_{2x} t_{2y}t_y^2}
\end{eqnarray}
As seen from (\ref{b11}), $R_\nu$ is proportional to $\cos \delta$ and 
therefore unless $\cos \delta$ is very small (i.e., $\delta$ is
very close to $\pi/2$) this pattern will not accommodate the observed 
data. Thus using $\delta=89^\circ$, we obtain
\begin{eqnarray}
&&\frac{m_1}{m_3} \; \approx \; \frac{m_2}{m_3} \; \approx 0.81
~~~~~R_\nu \approx 0.03\;,~~~~~~|M_{ee}|/m_3 \approx  0.81\;,
\nonumber\\
&& \frac{m_1}{m_3}-\frac{m_2}{m_3} \; \approx \; 0.007\; ,~~~ 
\sigma - \rho \; \approx \;
0.024\;.
\end{eqnarray}
\vspace{0.25cm}
Thus this pattern corresponds to the situation $m_1 \approx m_2 < m_3 $
and accommodates the observed data
on $R_\nu$ for $\delta$ close to $\pi/2$. 

\underline{Pattern $\rm B_2$:} ~ $M_{e\mu} = M_{\tau\tau} = 0$
(i.e., $a=b=\tau$; $\alpha =e$ and $\beta =\mu$). We obtain
\begin{eqnarray}
\frac{\lambda_1}{\lambda_3} & = & \frac{  s_x c_x c_y (2 s_z^2
s_y^2-c_y^2 c_z^2) + s_y s_z(c_x^2 s_y^2 e^{-i \delta }+c_y^2
s_x^2 e^{i \delta}) } { s_x c_x c_y s_y^2-(s_x^2 -c_x^2)s_y^3 s_z
e^{i \delta}+s_x c_x c_y s_z^2(1+s_y^2) e^{2 i \delta}} \; ,
\nonumber \\
\nonumber\\
\frac{\lambda_2}{\lambda_3} & = & \frac{ c_x s_x c_y (2 s_z^2
s_y^2-c_y^2 c_z^2)-s_y s_z(c_x^2 c_y^2 e^{i \delta}+s_x^2 s_y^2
e^{-i \delta})}{s_x c_x c_y s_y^2-(s_x^2 -c_x^2)s_y^3 s_z e^{i
\delta}+s_x c_x c_y s_z^2(1+s_y^2) e^{2 i \delta}} \; .
\end{eqnarray}
In the lowest-order approximation, we explicitly obtain
\begin{eqnarray}
&& \frac{m_1}{m_3} \; \approx \; \frac{m_2}{m_3} \; \approx \;
\frac{1}{t_y^2} \; ; ~~~ \rho \; \approx \; \sigma \; \approx  \pm
\frac{\pi}{2} \; ;
~~~~~
 R_\nu \; \approx \; \frac{1 + t^2_x}{t_x}~
 |t_{2y} ~ c_{\delta}| ~  s_z \; ,
\nonumber\\
&&|M_{ee}| \;  \approx  \; m_3 \left [ \frac{1}{t^2_y} -
\frac{c_\delta s_z}{t_x t_y} \left ( (1-t_x^2)(1+t_y^2)\right )
\right ] \; ,\nonumber\\
&& \frac{m_2}{m_3}-\frac{m_1}{m_3} \; \approx \; \frac{4 s_z
c_{\delta}}{s_{2y} s_{2x}} \; ,~~~ \sigma - \rho \; \approx \;
\frac{2 t_y^2 s_z s_{\delta}}{s_{2x} t_{2y}}\;.
\end{eqnarray}
As in B1, this case will also give acceptable solution for
$\delta$ close to $\pi/2$. Numerically the values are found for
$\delta=89^\circ$ as 
\begin{eqnarray}
&&\frac{m_1}{m_3} \; \approx \; \frac{m_2}{m_3} \; \approx1.23
~~~~~ R_\nu \approx 0.03\;,~~~~~~|M_{ee}|/m_3 \approx ~ 1.23\;,
\nonumber\\
&& \frac{m_1}{m_3}-\frac{m_2}{m_3} \; \approx \; 0.007\; ,~~~ 
\sigma - \rho \; \approx \;
0.02\;.
\end{eqnarray}
This pattern is similar to B1 and can accommodate the observed data
for $\delta$ close to $90^\circ$.This corresponds to the mass pattern
as $m_1 \approx m_2 >m_3$.

\underline{Pattern $\rm B_3$:} ~ $M_{\mu\mu} = M_{e\mu} = 0$
(i.e., $a=b=\mu$; $\alpha =e$ and $\beta =\mu$). We obtain
\begin{eqnarray}
\frac{\lambda_1}{\lambda_3} & = & \; - \frac{s_y}{c_y} ~.
\frac{s_x s_y-c_x c_y s_z e^{-i \delta}}{s_x c_y+c_x s_y s_z e^{i
\delta}}\; ,
\nonumber \\
\nonumber\\
\frac{\lambda_2}{\lambda_3} & =& \; - \frac{s_y}{c_y} ~. \frac{c_x
s_y + s_x c_y s_z e^{-i \delta}}{c_x c_y-s_x s_y s_z e^{i
\delta}}\; .
\end{eqnarray}
The approximate expressions for the neutrino mass ratios, the
Majorana phases and the observables $R_\nu$ and $|M_{ee}|$ turn out
to be
\begin{eqnarray}
&& \frac{m_1}{m_3} \; \approx \; \frac{m_2}{m_3} \; \approx \;
t^2_y \; ; ~~~ \rho \; \approx \; \sigma \; \approx  \pm
\frac{\pi}{2} \; ;
~~~~~~ R_\nu \; \approx \; \frac{1 + t^2_x}{t_x} t^2_y~
 |t_{2y} ~ c_{\delta}| ~  s_z \; ,
\nonumber\\
&&|M_{ee}| \;  \approx  \; m_3 \left [ t^2_y - \frac{c_\delta
s_z}{t_x t_y} \left ( (1-t_x^2)(1+t_y^2)t^2_y\right ) \right ] \;
,
\nonumber\\
&& \frac{m_2}{m_3}-\frac{m_1}{m_3} \; \approx \; \frac{4 s_z t^2_y
c_{\delta}}{s_{2y} s_{2x}} \; ,~~~ \rho - \sigma \; \approx \;
\frac{2 s_z s_{\delta}}{s_{2x} t_{2y}}\;.
\end{eqnarray}
The mass ratios are same as that of $B_3$ with $\delta=0$. However,
using $\delta=89^\circ$ the ratio of mass square difference 
$R_\nu$ and $|M_{ee}|$ are
found to be 
\begin{eqnarray}
R_\nu=0.025\;,~~~~|M_{ee}|/m_3=0.81\;,~~~~~ \frac{m_1}{m_3}-\frac{m_2}{m_3} 
\; \approx \; 0.005\; ,~~~ 
\rho - \sigma \; \approx \;
0.024\;.
\end{eqnarray}
Thus, this patten which was not acceptable for $\delta=0$ can accommodate the
observed data for $\delta=89^\circ$.
\vspace{0.25cm}

\underline{Pattern $\rm B_4$:} ~ $M_{\tau\tau} = M_{e\tau} = 0$
(i.e., $a=b=\tau$; $\alpha =e$ and $\beta =\tau$). We obtain
\begin{eqnarray}
\frac{\lambda_1}{\lambda_3} & = & \; - \frac{c_y}{s_y} \; .
\frac{s_x c_y+c_x s_y s_z e^{-i \delta}}{s_x s_y-c_x c_y s_z e^{i
\delta}}\; ,
\nonumber \\
\nonumber\\
\frac{\lambda_2}{\lambda_3} & =& \; - \frac{c_y}{s_y} \; .
\frac{c_x c_y - s_x s_y s_z e^{-i \delta}}{c_x s_y+s_x c_y s_z
e^{i \delta}}\; .
\end{eqnarray}
To lowest order, we get the following approximate results:
\begin{eqnarray}
&& \frac{m_1}{m_3} \; \approx \; \frac{m_2}{m_3} \; \approx \;
\frac{1}{t^2_y} \; ; ~~~ \rho \; \approx \; \sigma \; \approx  \pm
\frac{\pi}{2} \; ;
~~~~~ R_\nu \; \approx \; \frac{1 + t^2_x}{t_x t^2_y}~
 |t_{2y} ~ c_{\delta}| ~  s_z \; ,
\nonumber\\
&&|M_{ee}| \;  \approx  \; m_3 \left [ \frac{1}{t^2_y} +
\frac{c_\delta s_z}{t_x t_y} \left (
(1-t_x^2)(1+t_y^2)\frac{1}{t^2_y}\right ) \right ] \; ,
\nonumber\\
&& \frac{m_1}{m_3}-\frac{m_2}{m_3} \; \approx \; \frac{4 s_z
c_{\delta}}{t^2_y s_{2y} s_{2x}} \; ,~~~ \rho - \sigma \; \approx
\; \frac{2 s_z s_{\delta}}{s_{2x} t_{2y}}\;.
\end{eqnarray}
Thus, the numerical values of the mass parameters are given 
for $\delta=89^\circ$ as  
\begin{equation}
R_\nu \approx 0.04\;,~~~~~~|M_{ee}|/m_3 \approx 1.24\;,
~~~~~~
 \frac{m_1}{m_3}-\frac{m_2}{m_3} \; \approx \; 0.008\; ,~~~ 
\rho - \sigma \; \approx \;
0.024\;.
\end{equation}
Thus, the ratio of the square of mass difference is found to be ${\cal O}
(10^{-2})$ as observed by the current experiments.

\underline{Pattern $\rm C$:} ~ $M_{\mu\mu} = M_{\tau\tau} = 0$
(i.e., $a=b=\mu$; $\alpha =\beta =\tau$). We obtain
\begin{eqnarray}
\frac{\lambda_1}{\lambda_3} & = & \; - \frac{c_x c_z^2}{s_z} \; .
\frac{c_x(s_y^2-c_y^2)e^{-i \delta}+2 s_x s_y c_y s_z}{2 s_x c_x
s_y c_y-(s_x^2-c_x^2)(s_y^2-c_y^2)s_z e^{i \delta}+2 s_x c_x s_y
c_y s_z^2 e^{2i \delta}}\; ,
\nonumber \\
\nonumber\\
\frac{\lambda_2}{\lambda_3} & = &  \frac{s_x c_z^2}{s_z} \; .
\frac{s_x(s_y^2-c_y^2)e^{-i \delta}-2 c_x s_y c_y s_z}{2 s_x c_x
s_y c_y-(s_x^2-c_x^2)(s_y^2-c_y^2)s_z e^{i \delta}+2 s_x c_x s_y
c_y s_z^2 e^{2i \delta}}\; ,
\end{eqnarray}
To the lowest order , we get the mass ratios
\begin{eqnarray}
&& \frac{m_1}{m_3} \; \approx \; \sqrt{1-\frac{2 c_{\delta}}{t_x
t_{2y} s_z}+\frac{1}{t_x^2 t_{2y}^2 s_z^2}}
\; , ~~~~~~~\frac{m_2}{m_3} \; \approx \; \sqrt{1+\frac{2 t_x c_{\delta}}{
t_{2y} s_z}+\frac{t_x^2}{ t_{2y}^2 s_z^2}} \;,
\nonumber \\
&& \rho=\pm \frac{\pi}{2}+\frac{1}{2}\tan^{-1}\left (\frac{s_\delta}{t_x t_{2y}
s_z-c_\delta}
\right )\;,~~~~~ \sigma=\pm \frac{\pi}{2}-\frac{1}{2}\tan^{-1}\left (\frac{t_x
s_\delta}{t_{2y} s_z+t_xc_\delta}
\right )\;,\nonumber\\
&& R_\nu \; \approx \; \frac{1+t_x^2}{t_x^2}~
\left |\frac{2}{t_{2x}}~.\frac{1-t_{2x} t_{2y} c_{\delta} s_z}{t_x+2 s_z
c_{\delta} t_{2y}}\right |\;,
\nonumber\\
&&|M_{ee}| \;  \approx  \; m_3\sqrt{1-\frac{4 c_{\delta}}{t_{2x}
t_{2y} s_z}+\frac{4}{t_{2x}^2 t_{2y}^2 s_z^2}}\;.
\end{eqnarray}
In this case $R_\nu $ is very sensitive to $\delta$. Using
$\theta_z=5^\circ$ and $\delta=60^\circ$, we obtain the
numerical values of the mass parameters as
\begin{equation}
\frac{m_1}{m_3} \approx 1.55\;,~~~~
\frac{m_2}{m_3} \approx 1.57\;,~~~~R_\nu \approx 0.04\;,~~~~~~
|M_{ee}|/m_3 \approx ~ 0.99\;.
\end{equation} 
Thus, the pattern also gives acceptable solutions for $\delta=60^\circ$.
The order of magnitude of the 
mass matrix $M$ (4) for the seven possible forms for $\delta \neq 0$ are
presented in Table-2. 

To summarize, in this paper we have reanalyzed the seven possible 
forms of texture two neutrino mass matrices in the light of current
neutrino data. 
We found that, with standard parametrization, if the Dirac type 
CP violating phase in the neutrino
mixing matrix turns out to be zero, then out of the seven possible forms 
only two forms (A1, A2) are allowed by the current experimental data, which
corresponds to normal hierarchy. Furthermore, if we allow a slight variation
in $\theta_z$ then pattern C is also allowed but with inverted hierarchy.
For nonzero $\delta$ (Dirac CP phase) we have derived the expressions for the 
different mass ratios using the standard parametrization of the
neutrino mixing matrix, which are different from those
obtained in \cite{ref4}.
The mass matrices are also found to be slightly different.
Interestingly, when the  Dirac phase is nonzero all the possible
forms are allowed by the current data and most of the
structures (except $A_1$ and $A_2$, which are insensitive to the Dirac
phase and also follow normal hierarchy) 
do not exhibit normal hierarchy and prefer Dirac phase
close  to $\pi/2$.
In future, with more theoretical studies and with more
accurate data, we hope to understand better the true nature of 
the neutrino mass matrices.




\begin{table}
\caption{The allowed three patterns of the neutrino mass matrix 
$M$ with two texture
zeros, for $\delta=0$ in the mixing matrix.
The order-of-magnitude of $M$ is given for  illustration
for of $\theta_x =34^\circ$,~ $\theta_y=42^\circ$,~ $\theta_z=5^\circ$ for
A1, A2 and  $\theta_z=2.5^\circ$ for
pattern C.
}
\begin{center}
\begin{tabular}{ccccc} \hline\hline
Pattern &~~~& Texture of $M$ &~~~& Order of Magnitude \\ \hline
$\rm A_1$
&& $\left ( \matrix{
{\bf 0} & {\bf 0} & \times \cr
{\bf 0} & \times & \times \cr
\times & \times & \times \cr} \right )$
&&
$ \sim m_3 \left ( \matrix{
{\bf 0} ~ & ~ {\bf 0} ~ & ~ .12 \cr
{\bf 0} & .41 & .53 \cr
.12 & .53 & .51 \cr} \right )$
\\ \hline
$\rm A_2$
&& $\left ( \matrix{
{\bf 0} & \times & {\bf 0} \cr
\times & \times & \times \cr
{\bf 0} & \times & \times \cr} \right )$
&&
$\sim m_3 \left ( \matrix{
{\bf 0} ~ & ~ .13 ~ & ~ {\bf 0} \cr
.13 & .47 & .46 \cr
{\bf 0} & .46 & .59 \cr} \right )$
\\ \hline
$\rm C$
&& $\left ( \matrix{
\times & \times & \times \cr
\times & {\bf 0} & \times \cr
\times & \times & {\bf 0} \cr} \right )$
&&
$\sim m_3 \left ( \matrix{
.95 & 1.79 & 1.61 \cr
1.79 & {\bf 0} & 1.0 \cr
1.61 & 1.0 & {\bf 0} \cr} \right )$
\\ \hline\hline

\end{tabular}
\end{center}
\end{table}
\normalsize
\begin{table}
\caption{The allowed seven patterns of the neutrino mass matrix 
$M$ with two texture
zeros, for $\delta\neq0$ in the mixing matrix.
The order-of-magnitude of $M$ is given for  illustration
for of $\theta_x =34^\circ$,~ $\theta_y=42^\circ$,~ $\theta_z=5^\circ$,
$\delta=90^\circ$ for
A1 and A2, $\delta=89^\circ$ for B1, B2, B3, B4 and
$\delta=60^\circ$ for C.
}
\begin{center}
\begin{tabular}{ccccc} \hline\hline
Pattern &~~~& Texture of $M$ &~~~& Order of Magnitude \\ \hline
$\rm A_1$
&& $\left ( \matrix{
{\bf 0} & {\bf 0} & \times \cr
{\bf 0} & \times & \times \cr
\times & \times & \times \cr} \right )$
&&
$ \sim m_3 \left ( \matrix{
{\bf 0} ~ & ~ {\bf 0} ~ & ~ .12 \cr
{\bf 0} & .46 & .49 \cr
.12 & .49 & .54 \cr} \right )$
\\ \hline
$\rm A_2$
&& $\left ( \matrix{
{\bf 0} & \times & {\bf 0} \cr
\times & \times & \times \cr
{\bf 0} & \times & \times \cr} \right )$
&&
$\sim m_3 \left ( \matrix{
{\bf 0} ~ & ~ .13 ~ & ~ {\bf 0} \cr
.13 & .44 & .49 \cr
{\bf 0} & .49 & .56 \cr} \right )$
\\ \hline
$\rm B_1$
&& $\left ( \matrix{
\times & \times & {\bf 0} \cr
\times & {\bf 0} & \times \cr
{\bf 0} & \times & \times \cr} \right )$
&&
$\sim m_3 \left ( \matrix{
.81 & .02 & {\bf 0} \cr
.02 & {\bf 0} & .9 \cr
{\bf 0} & .9 & .19 \cr} \right )$
\\ \hline
$\rm B_2$
&& $\left ( \matrix{
\times & {\bf 0} & \times \cr
{\bf 0} & \times & \times \cr
\times & \times & {\bf 0} \cr} \right )$
&&
$\sim m_3 \left ( \matrix{
1.2 & {\bf 0} & .03 \cr
{\bf 0} & .23 & 1.1 \cr
.03 & 1.1 & {\bf 0} \cr} \right )$
\\ \hline
$\rm B_3$
&& $\left ( \matrix{
\times & {\bf 0} & \times \cr
{\bf 0} & {\bf 0} & \times \cr
\times & \times & \times \cr} \right )$
&&
$\sim m_3 \left ( \matrix{
.81 & {\bf 0} & .02 \cr
{\bf 0} & {\bf 0} & .90 \cr
.02 & .90 & .19 \cr} \right )$
\\ \hline
$\rm B_4$
&& $\left ( \matrix{
\times & \times & {\bf 0} \cr
\times & \times & \times \cr
{\bf 0} & \times & {\bf 0} \cr} \right )$
&&
$\sim m_3 \left ( \matrix{
1.2 & .03 & {\bf 0} \cr
.03 & .23 & 1.1 \cr
{\bf 0} & 1.1 & {\bf 0} \cr} \right )$
\\ \hline
$\rm C$
&& $\left ( \matrix{
\times & \times & \times \cr
\times & {\bf 0} & \times \cr
\times & \times & {\bf 0} \cr} \right )$
&&
$\sim m_3 \left ( \matrix{
.99 & .89 & .80 \cr
.89 & {\bf 0} & 1.0 \cr
.80 & 1.0 & {\bf 0} \cr} \right )$
\\ \hline\hline
\end{tabular}
\end{center}
\end{table}
\normalsize

\end{document}